\documentclass[twocolumn,showpacs,preprintnumbers,amsmath,amssymb,floatfix]{revtex4}

\usepackage{graphicx}
\usepackage{dcolumn}
\usepackage{bm}

\newcommand{\beq}{\begin{equation}}
\newcommand{\eeq}{\end{equation}}
\newcommand{\bey}{\begin{eqnarray}}
\newcommand{\eey}{\end{eqnarray}}

\begin{document}

\title{Central Density Dependent Anisotropic Compact Stars}

\author{Mehedi Kalam}
\email{kalam@iucaa.ernet.in} \affiliation{Department of
Physics, Aliah University, Sector - V, Salt Lake, Kolkata -
700091, West Bengal, India}

\author{Farook Rahaman}
\email{rahaman@iucaa.ernet.in} \affiliation{Department of
Mathematics, Jadavpur University, Kolkata 700032, West Bengal,
India}

\author{Sk. Monowar Hossein}
\email{sami_milu@yahoo.co.uk} \affiliation{Department of
Mathematics, Aliah University, Sector - V, Salt Lake, Kolkata -
700091, West Bengal, India}

\author{Saibal Ray}
\email{saibal@iucaa.ernet.in} \affiliation{Department of
Physics, Government College of Engineering \& Ceramic Technology, Kolkata -
700010, West Bengal, India}

\date{\today}

\begin{abstract}
Stars can be treated as self-gravitating fluid. In this
connection, we propose a model for an anisotropic
star under the relativistic framework of Krori-Barua \cite{Krori1975} 
spacetime. It is shown that the solutions are regular and singularity free. 
The uniqueness of the model is that interior physical properties of the star 
solely depend on the central density of the matter distribution.
\end{abstract}


\maketitle

\section{Introduction}
Late time evolution of stars due to strong gravity have been 
an interesting investigating field in astrophysics which let us know, 
via several physical processes, features of diverse characters 
of gravitating objects. So we actually come across a paradigm shift 
from the stage of normal stars to compact stars, ranging from 
dwarf stars to neutron stars, quark stars, dark stars, gravastars 
and eventually black holes. However, in the present study we are 
specifically interested for compact starnge stars category.

In connection to compact stars it have been shown by several workers 
\cite{Rahaman2012a,Rahaman2012b,Kalam2012a,Hossein2012,Kalam2012b} that the 
Krori and Barua \cite{Krori1975} (henceforth KB) spacetime provides an 
effective platform for modelling quark and strange type compact stars. 
Some of the interesting works to be mentioned with KB metric have been 
carried out by Rahaman et al. \cite{Rahaman2012a} for singularity-free 
dark energy stars which represents an anisotropic compact steller 
configuration with $8$~km radius whereas the same KB spacetime with 
MIT Bag model was considered by Rahaman et al. \cite{Rahaman2012b} 
to describe ultra-compact object like strange star with radius of $8.26$~km. 
Kalam et al. \cite{Kalam2012b} also proposed a model for strange quark stars 
within the framework of MIT Bag model. This model clearly indicates 
that the Bag constant need not necessarily lie within the range of 
$60 - 80~ MeV/fm^3$ as claimed in the literature.   

In the present work, we propose a model for an anisotropic compact star 
where it is assumed that the radial pressure exerted on the system is 
proportional to the matter density. The stellar configuration, therefore, comprises 
of two fluids - an ordinary baryonic matter together with an yet 
unknown form of matter (i.e. dark energy of Einstein-type varying cosmological 
constant) which is repulsive in nature. These two fluids are assumed to be 
non-interacting amongst themselves with the radial and transverse directional 
anisotropic property such that $p_r \neq p_t$ \cite{Rahaman2012a}. In favour of 
anisotropy we would like to point out that for fluid configuration this 
is not only natural but also obvious general option for describing relativistic 
compact stellar objects \cite{Bowers1974} (also see \cite{Herrera1992} 
for a review). Recent observational evidences on highly compact 
astrophysical objects like X-ray pulsar $Her X-1$, X-ray buster $4U 1820-30$, 
millisecond pulsar $SAX J 1808.4 - 3658$, X-ray sources $4U 1728 - 34$ etc. 
strongly favour an anisotropic pressure distributions in the 
case of an ultra-compact object. The mechanism at the microscopic level, 
though not yet well understood, may be related to variety of reasons such as 
the existence of type $IIIA$ superfluid, mixture of two fluids, phase transition, 
pion condensation, bosonic composition, rotation, magnetic field etc. 

As mentioned above that basically we have considered here a two fluids model 
for compact star with $\Lambda$-dark energy as one of the ingradients. 
This consideration has been motivated by the recent observations on the 
accelerated expansion of the universe. It is argued that about $73\%$ 
of the energy content of the universe is of gravitationally repulsive 
in nature ({\it dark energy}) \cite{Perlmutter1998,Riess2004}. As a result, 
cosmological models based on dark energy received much attention in the 
recent past either in the form of a non-zero cosmological constant 
(present value, $\Lambda \sim 1.1 \times 10^{-56} ~cm^{-2}$ \cite{Caldwell2009}) or in some other 
exotic forms of matter. However, in the astrophysical realm  dark energy may 
have implications not only with a pure cosmological constant 
\cite{Narlikar1991,Chen2009,Zubairi2010,Stuchlik2012} 
but also a varying cosmological constant \cite{Dymnikova2002,Egeland2007,Burdyuzha2009,MaK2000}. 
It has essentially been speculated that within a very massive dense star and/or galaxy 
a constant $\Lambda$ may have substantial effect and in the case of a space varying 
$\Lambda$ vary strengthwise from centre to boundary 
and decreases to extremly small non-zero value ($\Lambda \sim 10^{-56} ~cm^{-2}$) 
outside the system. In this connection we quote from Bambi \cite{Bambi2007} as he argues that 
``The cosmological constant problem represents an evident tension
between our present description of gravity and particle physics. Many solutions
have been proposed, but experimental tests are always difficult or impossible to
perform and the present phenomenological investigations focus only on possible
relations with the dark energy, that is with the accelerating expansion rate of
the contemporary universe. ... strange stars, if they exist,
could represent an interesting laboratory to investigate this puzzle, since their
equilibrium configuration is partially determined by the QCD vacuum energy
density''.

We would like to mention here the very recent work of Hossein et al. 
\cite{Hossein2012} where the authors have assumed cosmological constant 
with radial dependence i.e. $\Lambda ~=~ \Lambda(r)$ to describe non-singular
 model for anisotropic compact stars under the KB spacetime. 
In the present work we are essentially following the same root 
however with a quite different motivation. Our sole aim here is to find out 
a set of exact solution which is completely free from singularity (at $r=0$) 
and all the parameters involved in the solutions depend 
only on the central density ($\rho_0$) of the stellar interior. This then facilitates one 
to explore different physical parameters from a single parameter $\rho_0$. 

The plan of the present paper is as follows: In Sec II we have provided 
the basic equations in connection to the proposed model for strange star. 
Sec. III is dealt for finding out a class of exact solution for the steller interior of KB metric. 
Stability of the model is shown in Sec. IV whereas mass-radius relation 
and hence redshift calculation are done in Sec. V. Some concluding remarks 
are made in Sec. VI.

\section{Basic Formulations}
To describe the spacetime of a compact stellar configuration, 
we consider the metric of KB \cite{Krori1975} as given by
\begin{equation}
ds^2=-e^{\nu(r)}dt^2 + e^{\lambda(r)}dr^2 +r^2
(d\theta^2 +sin^2\theta d\phi^2), \label{eq1}
\end{equation}
with $\lambda(r)=Ar^2$ and $\nu(r) = Br^2 + C$. Here $A$, $B$ and $C$ are
 arbitrary constants to be determined on physical grounds.

We further assume that the energy-momentum tensor for the 
strange matter filling the interior of the star may be 
expressed in the following standard form as
\begin{equation}
T_{ij}=diag(\rho,-p_r,-p_t,-p_t),\label{eq2}
\end{equation}
where $\rho$, $p_r$ and $p_t$ correspond to respectively the energy density,
radial pressure and transverse pressure of the baryonic matter.

The Einstein field equations in the presence of $\Lambda$ can as usual be written as
\begin{equation}
 G_{ij} + \varLambda g_{ij} = -8\pi GT_{ij},\label{eq3}
\end{equation}
where the erstwhile cosmological constant $\Lambda$ is assumed to be 
space-dependent so that $\Lambda(r)~=~\Lambda_r~(say)$. Here $G = c=1$
under geometrized relativistic units.

\section{Interior structure}
In this model we propose that the radial pressure of 
a star is proportional to the matter density in the following form
\begin{equation}
p_r = \left(\frac{x}{1+x}\right)\rho,\label{eq4}
\end{equation}
where $x$ is any `+ve' real with the equation of state parameter 
$\omega(r)=x/(1+x)$. Here the choice is to ensure that $0 < \omega(r) < 1$.

By plugging the above expression for $p_r$ of Eq. (\ref{eq4}) into the Eq. (\ref{eq3}), we get 
\begin{eqnarray}
\rho =\frac{(1+x)(A+B)}{4\pi (1+2x)}e^{-Ar^2},\label{eq5}\\
p_{r} = \frac{x(A+B)}{4\pi (1+2x)}e^{-Ar^2},\label{eq6}\\
p_{t} = \frac{1}{8\pi (1+2x)}[e^{-Ar^2}(1+2x)(B^2
-AB)r^2 \nonumber \\ +(2Bx- A) - \frac{(1+2x)}{r^2}]+\frac{1}{8\pi r^2},\label{eq7}\\
\Lambda_r = \left[2(A\omega_r - B) - \frac{1+\omega_r}{r^2}\right]\frac{e^{-Ar^2}}{1+\omega_r} + \frac{1}{r^2}.\label{lambda}
\end{eqnarray}

Therefore, the equation of state parameters corresponding to the 
radial and transverse directions can be provided as
\begin{equation}
\omega_r(r) = \frac{x}{1+x},\label{eq8}
\end{equation}

\begin{eqnarray}
\omega_t(r) =
\frac{1}{2(1+x)(A+B)}\left[(B^2-AB)r^2(1+2x)\right. \nonumber\\
\left. +(2Bx-A)+A-\frac{1+2x}{r^2})+\frac{e^{Ar^2}(1+2x)}{r^2}\right].\label{eq9}
\end{eqnarray}

The measure of anisotropy, $\Delta = \left(p_{t}-p_{r}\right)$,
 in this model is obtained from the Eqs. (\ref{eq6}) and (\ref{eq7}) as follows:
\begin{multline}
\Delta  =
 \frac{e^{-Ar^2}}{8\pi }\left[\left((B^2 -AB)r^2  -A-\frac{1}{r^2}\right)+\frac{e^{Ar^2}}{r^2}\right].\label{eq10}
\end{multline}

To find out the expressions for the constants $A$ and $B$ of the KB model 
we match the interior metric to the exterior of the Schwarzschild solution
\begin{multline}
ds^{2}=-\left(1 - \frac{2M}{r} \right)dt^2
+ \left(1 - \frac{2M}{r} \right)^{-1}dr^2 \\+ r^2(d\theta^2+\sin^2\theta d\phi^2).
\label{eq11}
\end{multline}

Then assuming the boundary conditions smoothly on the surface 
of the stellar configuration at $r=R=$radius of the star, we get
\begin{eqnarray}
A &=& \frac{8 \pi \rho_{0}}{3},\label{eq12} \\
B &=& \frac{1}{2R^2} \left[ e^{\frac{8 \pi \rho_{0}}{3}
R^2}-1\right].\label{eq13}
\end{eqnarray}

At the center $r=0$, the density (\ref{eq7}) becomes 
\begin{equation}
 \rho_{0} = \frac{(1+x)(A+B)}{4\pi(1+2x)}.\label{eq14}
\end{equation}

Therefore, from the Eq. (\ref{eq14}), after substituting $\omega_r(r)$, $A$ and $B$ 
from the Eqs. (\ref{eq9}), (\ref{eq12}) and (\ref{eq13}), we get 
\begin{equation}
\omega_r(r)= \frac{1}{8\pi \rho_{0} R^2}\left[ e^{\frac{8 \pi \rho_{0}}{3}
R^2}-1 \right] -\frac{1}{3}.\label{eq15}
\end{equation}

This result immediately implies that the radial equation of state parameter 
of a particular star completely depends on the central density only.
It will then be convenient to rewrite the whole set of physical parameters from 
Eqs. (\ref{eq5}) - (\ref{eq7}) in terms of this central density as follows:
\begin{eqnarray}
\rho &=& \rho_{0}e^{-\frac{8 \pi \rho_{0}}{3}r^2},\label{eq16}\\
p_{r} &=& \frac{1}{8\pi R^2}\left[e^{\frac{8 \pi \rho_{0}}{3}(R^2-r^2)} 
- e^{-\frac{8 \pi \rho_{0}}{3}r^2}\right] \nonumber\\  & & -\frac{\rho_{0}}{3}e^{-\frac{8 \pi \rho_{0}}{3}r^2},\label{eq17}\\
p_{t} &=& \frac{e^{-Ar^2}}{8\pi}[(B^2
-AB)r^2 +\frac{2B(A+B-4\pi \rho_{0})}{A+B} \nonumber\\
& & - \frac{A(8\pi \rho_{0}-A-B)}{2(A+B)} - \frac{1}{r^2}]+\frac{1}{8\pi r^2},\label{eq18}
\end{eqnarray}
where $A$ and $B$ are already shown to be dependent on the central density 
as obvious from the Eqs. (\ref{eq12}) and (\ref{eq13}).

\begin{figure}[htbp]
    \centering
        \includegraphics[scale=.3]{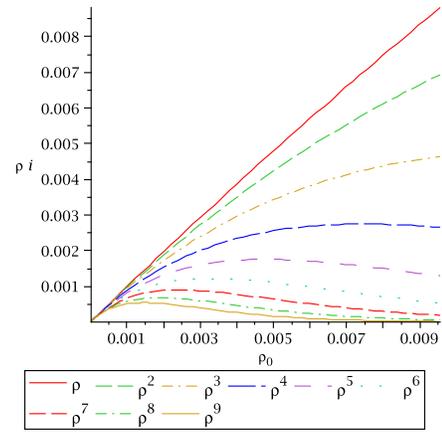}
        \caption{Density comparison of different star radii where 
$\rho^i$ indicate the density of the star of $`i'$~km radius.}
    \label{fig:1}
\end{figure}

\begin{figure}[htbp]
    \centering
        \includegraphics[scale=.5]{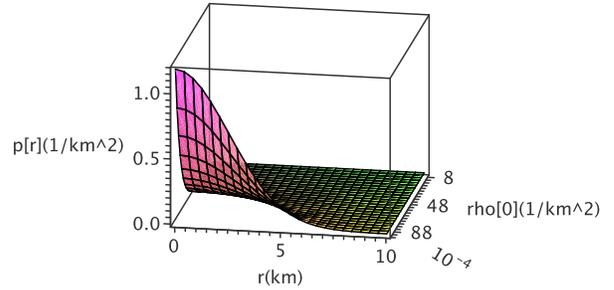}
        \caption{Pressure variation at the stellar interior.}
    \label{fig:2}
\end{figure}

From the above set of solution we observe the following two important features:

(1) Since the central density, $\rho_0$, being a non-zero constant quantity, 
all the physical parameters, viz. the matter density, radial pressure 
and transverse pressure of the star under consideration is entirely 
free from any singularity at $r=0$.

(2) At any distance from the center all the physical parameters 
completely depend only on the central density. In a similar way this 
also implies that the measure of anisotropy ($\varDelta$) 
and equation of state parameters ($\omega_{t}$ and $\omega_{t}$)
are central density dependent. Therefore, we can conclude taht at any distance 
from the center of the star, these are exactly measurable once 
we know the central density alone. 

These results are shown graphically in the Figs. 1 and 2. One can see that 
at $r=0$, the density becoms the central density $\rho_0$ itself. We see 
here that the central density of the star is low for comparatively bigger 
stars. Fig. 2 and 3 specially indicates that the present model is of 
compact strange star with radius about $10$~km. On the other hand the 
measure of anisotropy are plotted in the Figs. 3 and 4 to show the variation 
with respect to the central density. From these figures we can say that if 
the central density of the star becomes $<0.0068~km^{-2}$ then there should 
be no anisotropic behaviour within the stellar structure.

\begin{figure}[htbp]
    \centering
        \includegraphics[scale=.3]{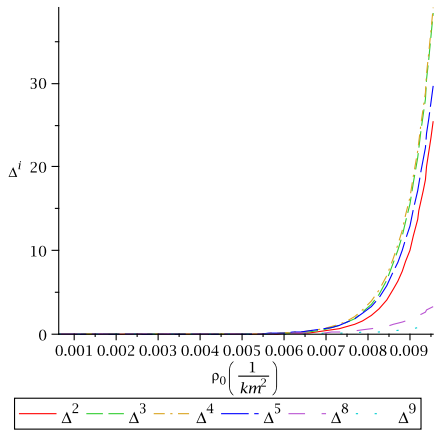}
        \caption{$\Delta^i$ variation with respect to central density 
at the stellar interior for different star radii where $\Delta^i$ indicate 
the anisotropy of the star of $`i'$~km radius.}
    \label{fig:3}
\end{figure}

\begin{figure}[htbp]
    \centering
        \includegraphics[scale=.4]{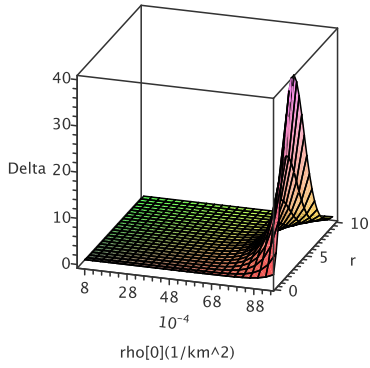}
        \caption{$\Delta$ variation with respect to central density at the stellar interior}
    \label{fig:4}
\end{figure}

\section{Stability}
For a physically acceptable model, one expects that the velocity
of sound should be within the range $0 \leq
v_s^2=(\frac{dp}{d\rho}) \leq 1$ \cite{Herrera1992,Abreu2007}. In
our anisotropic model, we define sound speeds as

\begin{equation}
v_{sr}^{2} ~=~ \omega_r(r) ~=~ \frac{x}{1+x}, \label{eq19}
\end{equation}

\begin{equation}
\begin{split}
v^2_{st}=\frac{dp_t}{d\rho}. \label{eq20}
\end{split}
\end{equation}

We plot the radial and transverse sound speeds in Fig. 5 
and observe that these parameters satisfy the inequalities $0\leq
v_{sr}^2 \leq 1$ and $0\leq v_{st}^2 \leq 1$ everywhere within
the stellar object.

\begin{figure}[htbp]
    \centering
        \includegraphics[scale=.3]{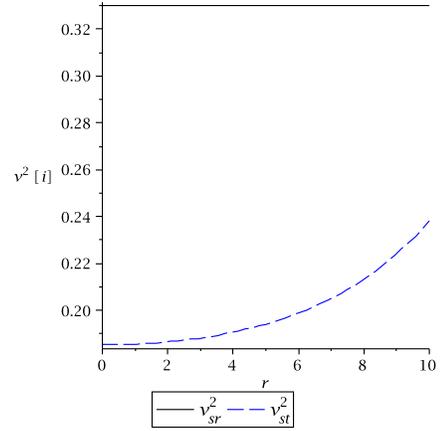}
        \caption{Sound speed variation of the stars.}
    \label{fig:5}
\end{figure}

\begin{figure}[htbp]
    \centering
        \includegraphics[scale=.5]{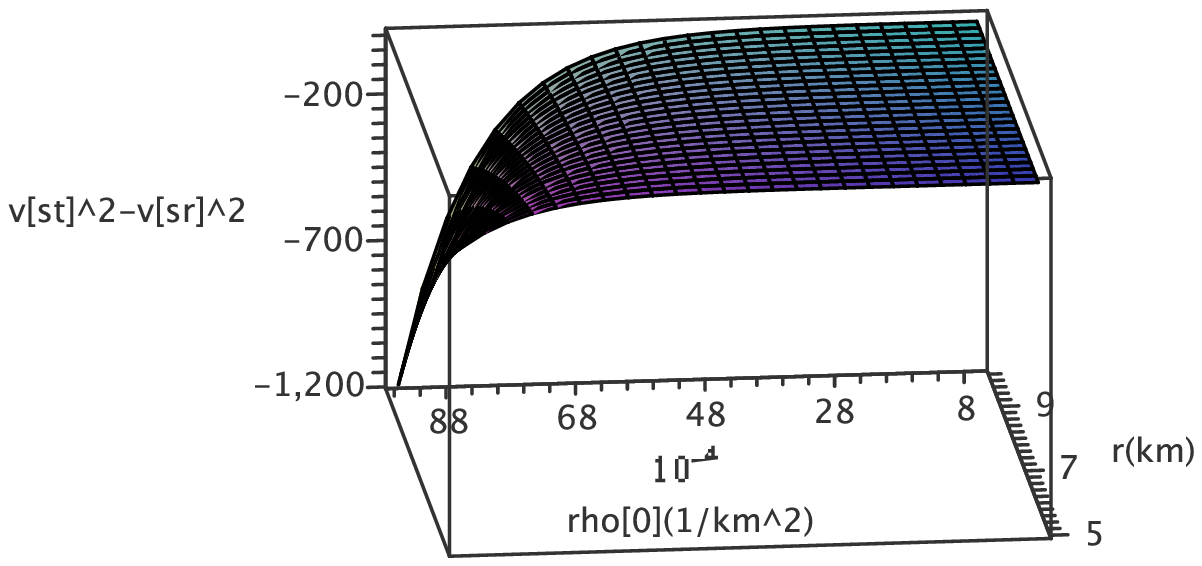}
        \caption{Sound speed variation of the stars with radii $>5$~km.}
    \label{fig:5}
\end{figure}

\begin{figure}[htbp]
    \centering
        \includegraphics[scale=.5]{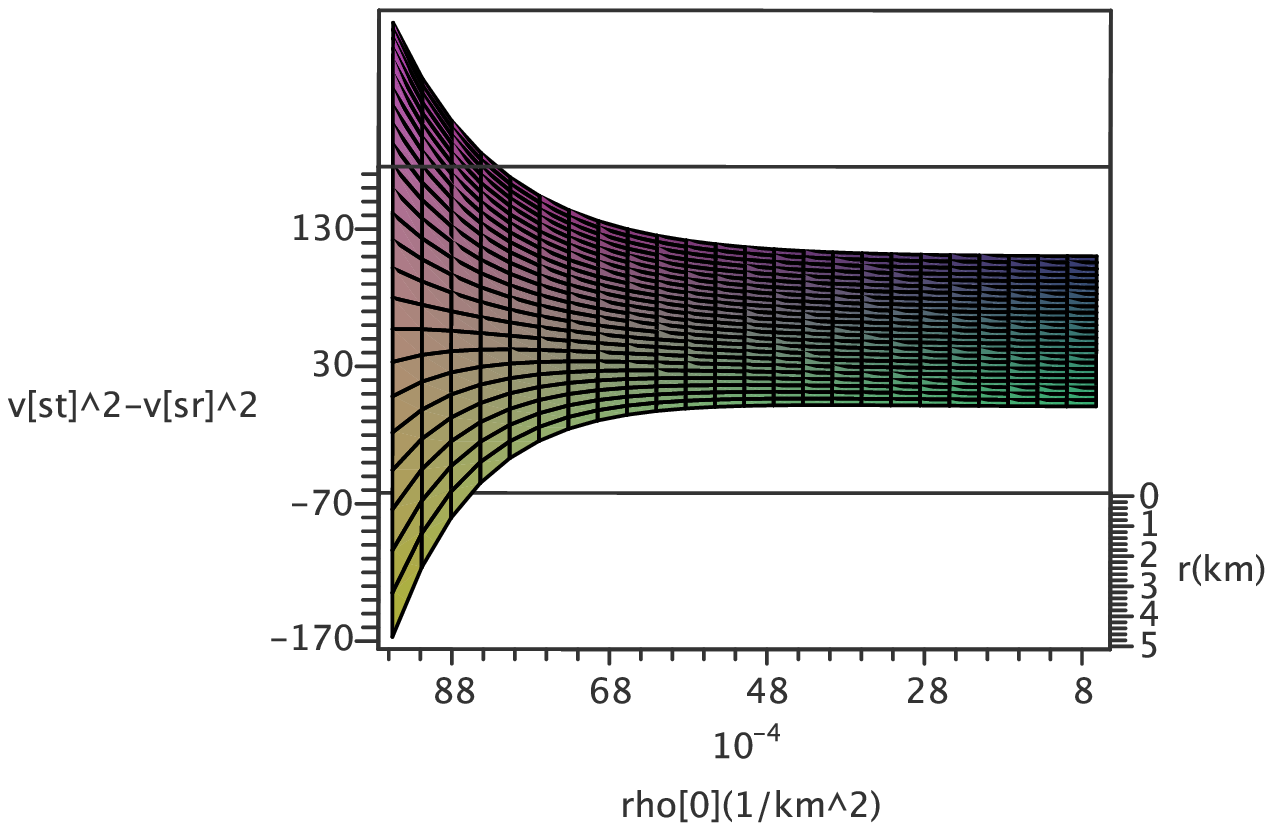}
        \caption{Sound speed variation of the stars with radii $\leq 5$~km.}
    \label{fig:6}
\end{figure}

Eqs. (\ref{eq19}) and (\ref{eq20}) lead to
\begin{widetext}
\begin{multline}
v^2_{st} - v^2_{sr}\\=\frac{(\omega_r+1)}{2A(A+B)}\left[
\frac{e^{Ar^2}}{r^4}+A(B^2-AB)r^2 -\left((B^2-AB)+\frac{A^2-A\omega_r(2B+A)}{(\omega_r+1)}\right) -
\frac{A}{r^2}-\frac{1}{r^4}\right]-\omega_r.\label{eq21}
\end{multline}
\end{widetext}

From the Eq. (\ref{eq21}) we note that $|v^2_{st}-v^2_{sr}|\leq 1$. 
Since, $0\leq v_{sr}^2 \leq 1$ and $0\leq v_{st}^2 \leq 1$,
therefore, $\mid v_{st}^2 - v_{sr}^2 \mid \leq 1 $.

Now, to examine the stability of local anisotropic matter
distribution, we use Herrera's \cite{Herrera1992} {\em cracking} 
(also known as {\em overturning}) concept which states that the 
region for which radial speed of sound is greater than the transverse speed 
of sound is a potentially stable region. In our case, Fig. 6 indicates that 
there is no change of sign for the term $v_{st}^2 - v_{sr}^2$ within the 
specific region of the configuration. This implies that the KB-type strange star
model is a stable one if the radius of the star is $>5$~km. From the Fig. 7  
it is evident that if radius of a star is $<5$~km 
then the central density should be $>68 \times 10^{-4}~km^{-2}$ [approx] 
to be stable according to our proposed model.

\section{Surface Redshift}
In this section, we study the maximum allowable mass-radius ratio
in our model. For a static spherically symmetric perfect fluid
star, Buchdahl \cite{Buchdahl1959} showed that the maximally
allowable mass-radius ratio is given by $\frac{2M}{R} <
\frac{8}{9}$ (for a more generalized expression for the same see
Ref. \cite{Mak2001}). 

Now we write the compactness of the star in the following form:

\begin{eqnarray}
u= \frac{M}{R} ~~~~~~~~~~~~~~~~~~~~~~~~~~~~~~~~~~~~~~~~~~~~~~~~~~~~ \nonumber\\ 
= - \frac{(A+B)}{4R(\omega_r+1)A^{\frac{5}{2}}}\left[2Re^{-AR^2}A^{\frac{3}{2}}+
\sqrt{\pi}~ erf(\sqrt{A}R)A\right].\label{eq22}
\end{eqnarray}

The surface redshift ($Z_s$) corresponding to the above
compactness factor ($u$) is obtained as

\begin{widetext}
\begin{eqnarray}
Z_s= [1-2u]^{-\frac{1}{2}} - 1  ~~~~~~~~~~~~~~~~~~~~~~~~~~~~~~~~~~~~~~~~~~~~~~~~~ \nonumber\\
= \left[1+\frac{(A+B)}{2R(\omega_r+1)A^{\frac{5}{2}}}\left(2Re^{-AR^2}A^{\frac{3}{2}}+
\sqrt{\pi}~ erf(\sqrt{A}R)A\right)\right]^{-\frac{1}{2}} - 1. \label{eq24}
\end{eqnarray}
\end{widetext}

It is clear from the above expression that redshift is 
dependent on $\rho_{0}$ and also it is well behaved (Fig 8).

\begin{figure}[htbp]
    \centering
        \includegraphics[scale=.3]{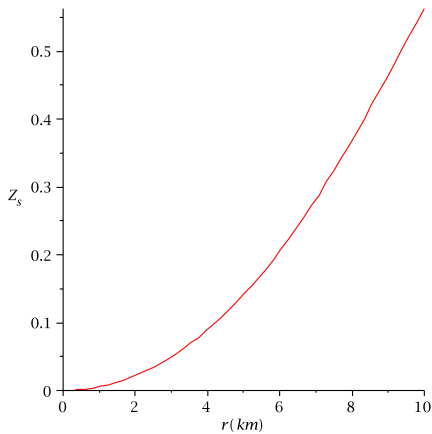}
        \caption{Redshift variation at the stellar interior.}
    \label{fig:7}
\end{figure}

\section{Conclusion}
Keeping in mind that stars can be treated as self-gravitating fluid 
we have proposed a very simple and unique model for an anisotropic star. 
The spacetime here is of KB-type \cite{Krori1975}
which is supposed to present compact strange stars solutions 
\cite{Rahaman2012a,Rahaman2012b,Kalam2012a,Hossein2012,Kalam2012b}. 
It is shown through the solution set that (1) the central density 
$\rho_0$ being a non-zero constant quantity the star is non-singular 
at $r=0$, and (2) interior physical properties 
of a compact star solely depend on the central density of the 
stable steller configuration. Thus if we know 
the {\em mass-radius ratio} then all the interior features of 
a star can be exactly evaluated at any position of the 
interior of that star. 

Further it may be observed that if 
we know the central density and radius of a star then 
assuming the spherical structure we can say that matter 
distribution follow a regular behavioural pattern whatever the 
size of a star may be. We also show that the central density 
is proportoinately low for bigger star.

As a final comment, following the introductory discussion regarding 
inclusion of cosmological constant in the source term 
of Einstein field equations as one of the candidates of dark energy, 
we argue that ``Strange stars, if they exist, can play 
an important role in the solution to the cosmological constant problem''\cite{Bambi2007}.

\vspace{1.0cm}

\section*{Acknowledgments} MK, FR and SR gratefully acknowledge 
support from IUCAA, Pune, India under Visiting Associateship Programme 
under which a part of this work was carried out. SMH is also 
thankful to IUCAA for giving him an opportunity to visit IUCAA 
where a part of this work was carried out. FR is personally thankful 
to PURSE, DST and UGC for providing financial support.

\end{document}